\documentclass[aps,twocolumn,epsfig,prb]{revtex4}

\newcommand{\bepsilon}{\mbox{\boldmath $\epsilon$}}
\newcommand{\bv}{\mbox{\boldmath $v$}}

\usepackage{amsmath}
\usepackage{graphicx}

\def\nn{\nonumber}

\begin{document}

\title{Theory of optical transitions in graphene nanoribbons}

\author{Ken-ichi Sasaki}
\email[Email address: ]{sasaki.kenichi@lab.ntt.co.jp}
\affiliation{NTT Basic Research Laboratories, 
Nippon Telegraph and Telephone Corporation,
3-1 Morinosato Wakamiya, Atsugi, Kanagawa 243-0198, Japan}

\author{Keiko Kato}
\affiliation{NTT Basic Research Laboratories, 
Nippon Telegraph and Telephone Corporation,
3-1 Morinosato Wakamiya, Atsugi, Kanagawa 243-0198, Japan}

\author{Yasuhiro Tokura}
\affiliation{NTT Basic Research Laboratories, 
Nippon Telegraph and Telephone Corporation,
3-1 Morinosato Wakamiya, Atsugi, Kanagawa 243-0198, Japan}

\author{Katsuya Oguri}
\affiliation{NTT Basic Research Laboratories, 
Nippon Telegraph and Telephone Corporation,
3-1 Morinosato Wakamiya, Atsugi, Kanagawa 243-0198, Japan}

\author{Tetsuomi Sogawa}
\affiliation{NTT Basic Research Laboratories, 
Nippon Telegraph and Telephone Corporation,
3-1 Morinosato Wakamiya, Atsugi, Kanagawa 243-0198, Japan}

\date{\today}

\begin{abstract}
 Matrix elements of electron-light interactions
 for armchair and zigzag graphene nanoribbons 
 are constructed analytically using a tight-binding model.
 The changes in wavenumber ($\Delta n$)
 and pseudospin are the necessary elements 
 if we are to understand the optical selection rule. 
 It is shown that 
 incident light with a specific polarization and energy, 
 induces an indirect transition ($\Delta n=\pm1$),
 which results in a characteristic peak in the absorption spectra.
 Such a peak provides evidence that
 the electron standing wave is formed by multiple reflections at both
 edges of a ribbon. 
 It is also suggested that 
 the absorption of low-energy light is sensitive to the position of the
 Fermi energy, direction of light polarization, and irregularities in
 the edge. 
 The effect of depolarization on the absorption peak is briefly discussed.
\end{abstract}

\maketitle

\section{Introduction}

The dynamics of the electrons in graphene
is governed by an equation 
that is similar to the relativistic equation for the
massless Dirac fermion.~\cite{novoselov05,zhang05}
Since the Fermi velocity of a carrier
is about $10^6$m/s, 
the Dirac fermion moves forward 1 nm in 1 fs.
Therefore, in nanometer-sized graphene, 
the carrier reaches the edge before 
its motion is affected by perturbations such as 
electron-phonon and electron-electron interactions.
As a result, the electronic properties of the system 
are sensitive to the presence of an edge.
The behavior of electrons
near the edge of a graphene sample is unique
due to the reflection of the massless Dirac fermion.

In graphene nanoribbons,
the importance of the edge is marked by reflections of electron
taking place at the both edges of the ribbon.~\cite{li08,jia09,wang10,berry87}
The reflections result in the formation of 
a standing wave of a Dirac fermion.
In this paper, we examine optical transitions
in graphene nanoribbons and clarify characteristic features of the
standing wave of a Dirac fermion.
Knowing the rules for the optical transitions 
is an important step in understanding the optical properties
of graphene nanoribbons.


Graphene edges are categorized into two groups:
armchair and zigzag edges 
with respect to the symmetry of the hexagonal lattice.~\cite{tanaka87,fujita96,nakada96}
It is known that 
standing waves near the armchair edge 
and near the zigzag edge
are distinct for various reasons,
such as their pseudospin and Berry's phase.~\cite{sasaki10-forward} 
In this paper, 
we study nanoribbons with armchair and zigzag edges in great detail, 
and briefly discuss the effect of irregularities 
in the edge on the absorption spectra.

Here, we mention previously published literature on 
the optical absorption of graphene nanoribbons.
Hsu and Reichl~\cite{hsu07} 
investigated the absorption of linearly polarized light 
parallel to a zigzag nanoribbon 
and found that a direct transition is not allowed,
in contrast to the case of nanotubes.~\cite{ajiki94}
Gundra and Shukla~\cite{gundra11} pointed out that 
the polarization dependence on the absorption spectra 
is important for characterizing the edge structure.
Whereas these studies were based on numerical simulations, 
our study provides analytical results for electron-light matrix elements. 
The analytical result clearly shows
the effects of pseudospin and momentum conservation 
on the optical properties of a graphene nanoribbon.

This paper is organized as follows. 
In Sec.~\ref{sec:armrib},
we deal with armchair nanoribbons.
By constructing the matrix elements of the electron-light interaction,
we show that dynamical conductivity 
depends on the direction of the polarization 
of the incident light with respect to the orientation of the edge.
In Sec.~\ref{sec:zigrib}, we examine absorption spectra for zigzag
nanoribbons. The effect of edge irregularities
on dynamical conductivity is studied in Sec.~\ref{sec:mixed}.
Our discussion and conclusion are provided in Secs.~\ref{sec:dis}
and~\ref{sec:con}, respectively.

\section{Armchair nanoribbon}\label{sec:armrib}

In this section 
we study the electron-light interaction in armchair nanoribbons.
In Sec.~\ref{ssec:armwf}, 
we review the electronic properties of armchair nanoribbons
to provide the necessary background.
In Sec.~\ref{ssec:optpol},
the matrix elements of electron-light interaction are constructed.
The results are used to clarify the absorption spectra 
in Sec.~\ref{ssec:dc}.

\subsection{Electron wavefunction}\label{ssec:armwf}

\begin{figure}[htbp]
 \begin{center}
  \includegraphics[scale=0.45]{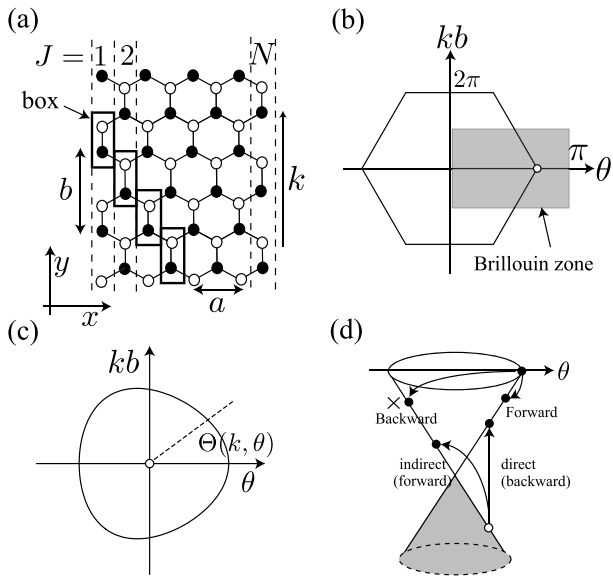}
 \end{center}
 \caption{ (a) The structure of an $N$ armchair nanoribbon.
 The unit length along the edge is denoted by $b$ ($\equiv \sqrt{3}a$).
 Carbon atoms are divided into A ($\bullet$) and B ($\circ$) atoms. 
 (b) The BZ of armchair nanoribbons: 
 $\theta\in(0,\pi)$ and $kb \in[-\pi,\pi)$.
 The circle stands for the Dirac point.
 (c) The definition of the polar angle $\Theta(k,\theta)$.
 The circle represents an energy contour (1 eV). 
 (d) Schematic diagrams for intra- and inter-band transitions from the
 initial state with $\Theta_n=0$ (on the $\theta$-axis).
 }
 \label{fig:arm}
\end{figure}

The energy dispersion relation for armchair nanoribbons~\cite{compernolle03,zheng07,sasaki11-armwf}
is written as 
\begin{align}
 \varepsilon^s(k,\theta)=s\gamma_0 
 \sqrt{1+4\cos^2\theta+4\cos\theta\cos kl},
 \label{eq:enearm}
\end{align} 
where $\gamma_0$ is the hopping integral ($\gamma_0=3$ eV) 
and $l\equiv \sqrt{3}a/2$ ($a$ is a lattice constant [$a=2.46$ \AA]).
The energy is characterized 
by the band index $s$ and two parameters $k$, and $\theta$.
The superscript, $s$, represents the conduction (valence) energy band and 
takes values $s=1$ ($-1$) on the right-hand side of Eq.~(\ref{eq:enearm}),
$k$ is the wave vector parallel to the edge, 
and $\theta$ stands for the phase in the direction perpendicular to the
edge [see Fig.~\ref{fig:arm}(a)].
By setting $k=0$ and $\theta=2\pi/3$ into Eq.~(\ref{eq:enearm}),
we see that $\varepsilon^s(0,2\pi/3)=0$ is satisfied.
Thus, the conduction and valence energy bands touch at
the point $(k,\theta)=(0,2\pi/3)$, which is called the Dirac point.
The energy dispersion relation for armchair nanoribbons
Eq.~(\ref{eq:enearm}) is the same as that for graphene.~\cite{saito98book}
There are two independent Dirac points
(known as K and K$'$ points) in the Brillouin zone (BZ) of graphene, 
on the other hand, 
there is a single Dirac point in the BZ of armchair nanoribbons.
That is, although $\varepsilon^s(k,\theta)$ is zero at the other point
$(k,\theta)=(0,-2\pi/3)$, this point is not included in the BZ of
armchair nanoribbons.
In fact, the BZ of armchair nanoribbons
is given by $\theta \in (0,\pi)$
and $kb \in [-\pi,\pi)$, where $b\equiv \sqrt{3}a$,
as shown in Fig.~\ref{fig:arm}(b).
The BZ of armchair nanoribbons 
covers only one-half of the graphene's BZ
because the reflection taking place at the armchair edge
identifies $\theta$ with $-\theta$.~\cite{sasaki11-armwf}

The wave function is given by
\begin{align}
 \phi^s_J(k,\theta) = \frac{1}{\sqrt{N}}e^{-ikl(J-1)} \sin J\theta
 \begin{pmatrix}
  e^{-i\Theta(k,\theta)} \cr s 
 \end{pmatrix},
 \label{eq:bw}
\end{align}
where $J$ $(=1,\ldots,N)$ is the coordinate 
perpendicular to the edge [see
Fig.~\ref{fig:arm}(a)].~\cite{compernolle03,zheng07,sasaki11-armwf} 
A detailed derivation of the wave function 
is given in Ref.~\onlinecite{sasaki11-armwf}.
The upper (lower) component of Eq.~(\ref{eq:bw})
represents the amplitude at A-atom (B-atom)
in the box shown in Fig.~\ref{fig:arm}(a).
The relative phase between the two components,
$\Theta(k,\theta)$ in Eq.~(\ref{eq:bw}),
is the polar angle defined with respect
to the Dirac point as shown in Fig.~\ref{fig:arm}(c).
The wave function with $\theta = 2\pi/3$
does not have an amplitude at $J=3,6,\ldots$ 
due to $\sin(J\theta)$ of Eq.~(\ref{eq:bw}).
The corresponding line nodes have been observed
in recent scanning tunneling microscopy topography,~\cite{yang10}
which evidences the wave function Eq.~(\ref{eq:bw}).

The wave function vanishes at $J=0$ and $J=N+1$ 
(i.e., at fictitious edge sites).
The boundary condition for $J=0$ 
is given by $\phi^s_{J=0}(k,\theta)=0$, which is 
satisfied with arbitrary values of $\theta$.
The phase $\theta$ is quantized 
by the boundary condition for $J=N+1$, 
$\phi^s_{N+1}(k,\theta)=0$, as 
\begin{align}
 \theta_n = \frac{n\pi}{N+1}, \ \ (n=1,\ldots,N),
 \label{eq:theta}
\end{align}
where $n$ represents the subband index.
Meanwhile, we assume that the wave vector $k$ 
parallel to the edge is a continuous variable. 
Because $k$ is not changed by the perturbations,
which preserve translational symmetry along the edge, 
hereafter, we abbreviate 
$\phi^s_J(k,\theta_n)$, $\Theta(k,\theta_n)$, and $\varepsilon^s(k,\theta_n)$
by omitting $k$ and $\theta$
as $\phi^s_{n,J}$, $\Theta_n$, and $\varepsilon^s_n$, respectively.

\subsection{Selection rule}\label{ssec:optpol}

The electron-light interaction is written as 
$H_{\rm em}=-e \bv\cdot {\bf A}$,
where $-e$ is the electron charge,
$\bv=(v_x,v_y)$ is the velocity operator,
and ${\bf A}\propto e^{-i\omega t}\bepsilon$ is a spatially
uniform vector potential.
Here, $\hbar \omega$ corresponds to the energy of the incident light
and $\bepsilon$ denotes the polarization.
The matrix elements of $\bv$
are classified into inter-band
$\langle \phi_n^c | \bv | \phi_m^v \rangle$
and intra-band $\langle \phi_n^c | \bv | \phi_m^c \rangle$ transitions.
The contributions of these transitions to the optical property
of a nanoribbon depend not only on $\hbar \omega$ and $\bepsilon$, but
also on the position of the Fermi energy $\varepsilon_{\rm F}$.
The intra-band transition may be omitted only when 
$\varepsilon_{\rm F}=0$ (band center) at zero temperature.
In general, it is necessary to consider both transitions.

First, we consider the intra-band transition.
The matrix elements of $\bv$
are calculated using Eq.~(\ref{eq:bw}).
The details of the calculation are provided in Appendix~\ref{app:v}.
The results are
\begin{align}
 & 
 \langle \phi^c_m | v_x | \phi^c_n \rangle 
 = \begin{cases}
  \displaystyle 0 & \text{$m-n \in$ even}\\
  \displaystyle -i \frac{2}{\pi} \frac{v_{\rm F}}{m-n} \langle \sigma_x
    \rangle_{mn} & \text{$m-n \in$ odd}, 
 \end{cases}
 \label{eq:ve1} \\
 & 
 \langle \phi^c_m | v_y | \phi^c_n \rangle 
 = \delta_{mn}v_{\rm F} \langle \sigma_y \rangle_{mn},
 \label{eq:ve}
\end{align}
where $\delta_{mn}$ is the Kronecker delta, $v_{\rm F}$ 
($\equiv \gamma_0 l/\hbar$) is the Fermi velocity, and
$\sigma_i$ ($i=x,y$) denotes the Pauli matrices.
The matrix elements of $\sigma_i$ are written as
\begin{align}
 & \langle \sigma_x \rangle_{mn}
 = \frac{1}{2} \left( e^{i\Theta_m} + e^{-i\Theta_n}\right), 
 \label{eq:sigx} \\
 & \langle \sigma_y \rangle_{mn}
 = -\frac{i}{2} \left( e^{i\Theta_m} - e^{-i\Theta_n} \right).
 \label{eq:sig}
\end{align}
$\sigma_i$ is the pseudospin, 
which brings certain features to the optical transitions.
For example, 
in Eq.~(\ref{eq:sigx}),
$\langle \sigma_x \rangle_{mn}=1$ is satisfied 
for forward scattering from $\Theta_n =0$
to $\Theta_{m}=0$, while 
$\langle \sigma_x \rangle_{mn}=0$ for 
backward scattering from $\Theta_n =0$ to $\Theta_m=\pi$ 
[see Fig.~\ref{fig:arm}(d)].
Since $\langle \sigma_x \rangle_{mn}$
is proportional to $\langle \phi^c_m | v_x | \phi^c_n \rangle$ 
as described in Eq.~(\ref{eq:ve1}),
the $x$-polarized light ($A_x$) does not cause
intra-band backward scattering.

In Eq.~(\ref{eq:ve1}),
the factor in front of the pseudospin $\langle \sigma_x \rangle_{mn}$
arises from momentum conservation.
It can be shown that momentum conservation 
for $\langle \phi^c_m | v_x | \phi^c_n \rangle$ 
leads to the following summation with respect to 
the out-of-phase trigonometric functions (Appendix~\ref{app:v}):
\begin{align}
 \frac{2}{N}\sum_{J=1}^N \sin (J\theta_m) \cos(J\theta_n) =
 \begin{cases}
  \displaystyle 0 & \text{$m-n \in$ even} \\
  \displaystyle 
  \frac{2}{\pi} \frac{1}{m-n}
  & \text{$m- n \in$ odd},
 \end{cases}
 \label{eq:momsum}
\end{align}
where $\theta_{m(n)}$ is given by Eq.~(\ref{eq:theta}).
The summation takes a non-zero value only when
$m-n$ ($\equiv \Delta n$) is an odd number.
The momentum conservation for $\langle \phi^c_m | v_y | \phi^c_n \rangle$ 
leads to the summation of the in-phase 
trigonometric functions [see Eq.~(\ref{app:inphase})], 
so that the summation takes a non-zero value only for $m=n$.
Hereafter, 
we call the process satisfying $m-n \in$ 
odd an indirect transition for which
the wavenumber of the initial state changes ($\Delta n \in$ odd), 
and the process satisfying $m=n$ is called a direct transition ($\Delta n=0$).

We have seen that 
the velocity matrix element is determined 
by two factors:
pseudospin and momentum conservation.
Next, 
we consider the inter-band transition based on this understanding.
The calculated matrix elements are given by 
\begin{align}
 & \langle \phi^c_m | v_x | \phi^v_n \rangle 
 = \begin{cases}
  \displaystyle 0 & \text{$m-n \in$ even}\\
  \displaystyle - \frac{2}{\pi} \frac{v_{\rm F}}{m-n} \langle \sigma_y \rangle_{mn} & \text{$m-n \in$ odd},
 \end{cases}
\label{eq:interx} \\
 & \langle \phi^c_m | v_y | \phi^v_n \rangle 
 = i \delta_{mn} v_{\rm F} \langle \sigma_x \rangle_{mn}.
\label{eq:intery}
\end{align}
In Eq.~(\ref{eq:intery}),
$\delta_{mn}$ shows that the $y$-polarized light ($A_y$)
results in a direct inter-band transition
[see Fig.~\ref{fig:arm}(d)].
Thus, the transition amplitude depends on 
the diagonal matrix element
of the pseudospin $\langle \sigma_x \rangle_{nn}$.
From Eq.~(\ref{eq:sigx}), 
we see that 
$\langle \sigma_x \rangle_{nn}$ takes a maximum value of  
$\langle \sigma_x \rangle_{nn}=\cos\Theta_n = \pm 1$
for $\Theta_n=0$ or $\pi$.
On the other hand, 
$\langle \sigma_x \rangle_{nn}$ vanishes for 
$\Theta_n=\pm \pi/2$.
Therefore,
the electrons on the $\theta$-axis ($\Theta_n=0$ or $\pi$) 
are selectively excited by $A_y$, 
while the electrons near the $k$-axis ($\Theta_n=\pm \pi/2$)
are excited very little.

As we can see in Eq.~(\ref{eq:interx}),
the $x$-polarized light ($A_x$)
results in an indirect transition 
[see Fig.~\ref{fig:arm}(d)].
An inter-band transition amplitude 
from $\Theta_n=0$ to $\Theta_m=\pi$
is enhanced because the strength of the pseudospin takes a maximum value
of $|\langle \sigma_y \rangle_{mn}|=1$. 
This indirect inter-band transition
is a forward scattering, namely, 
the electron crosses the Dirac point.
An inter-band transition 
that is not across the Dirac point, such as 
a (backward) transition from $\Theta_n=0$ to $\Theta_m=0$,
is allowed by momentum conservation if $m-n$ is an odd number. 
However, it is strongly suppressed by the pseudospin
because $\langle \sigma_y \rangle_{mn} = 0$
for the process.

\subsection{Dynamical conductivity}\label{ssec:dc}

With the optical selection rule established by
Eqs.~(\ref{eq:ve1}), (\ref{eq:ve}), (\ref{eq:interx}), and (\ref{eq:intery}), 
let us investigate the absorption of light.
The dynamical conductivity is given by ($\alpha=x,y$)
\begin{align}
 \sigma_{\alpha\alpha}(\omega) = 
 \frac{\hbar}{iS} \sum_{s's}\sum_{nmk} 
 \frac{\left(f[\varepsilon^{s'}_{m}]-f[\varepsilon^s_{n}]\right)
 \left| \langle \phi^{s'}_m | (-e v_\alpha) | \phi^{s}_{n} \rangle
 \right|^2}{\left(\varepsilon^{s'}_m-\varepsilon^s_{n}\right)
 \left(\varepsilon^{s'}_m-\varepsilon^s_{n}+ \hbar \omega +i\delta
 \right)},
 \label{eq:dcformula}
\end{align}
where $f[\varepsilon]$ is the Fermi distribution function,
$S$ is the nanoribbon area, and 
$\delta$ is inversely proportional to the
relaxation time of the excited electron.~\cite{ando02-dc}
The real part of the dynamical conductivity,
${\rm Re}(\sigma_{xx})$ (${\rm Re}(\sigma_{yy})$), 
represents the absorption spectrum of $x$-polarized 
($y$-polarized) light.
In Eq.~(\ref{eq:dcformula}), 
we assume $\delta=10$ meV
and room temperature when evaluating the Fermi distribution function.

\begin{figure}[htbp]
 \begin{center}
  \includegraphics[scale=0.4]{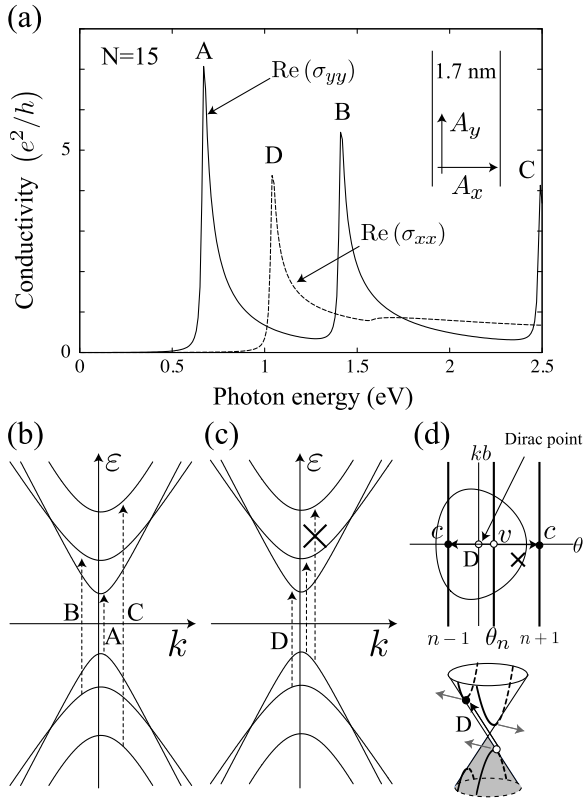}
 \end{center}
 \caption{ (a) The calculated ${\rm Re}(\sigma_{\alpha\alpha})$
 for an $N=15$ (semiconducting) armchair nanoribbon.
 The inter-band transition processes relevant to the absorption peaks of
 ${\rm Re}(\sigma_{yy})$ and ${\rm Re}(\sigma_{xx})$ 
 are shown in (b) and (c), respectively.
 (d) The optically allowed indirect inter-band transition 
 (from $\phi_n^v$ to $\phi_{n-1}^c$ or from $\phi_{n-1}^v$ to
 $\phi_{n}^c$).
 Bottom: The pseudospin (denoted by arrows) 
 is useful for showing that only forward scattering is allowed in the
 inter-band optical transitions. 
 }
 \label{fig:dc_arm}
\end{figure}

The solid curve in Fig.~\ref{fig:dc_arm}(a) 
shows ${\rm Re}(\sigma_{yy})$
calculated for an $N=15$ armchair nanoribbon.
The width of this nanoribbon is 1.7 nm.
As we have shown in Eq.~(\ref{eq:intery}), 
the absorption of $y$-polarized light 
results from a direct inter-band transition.
Thus, the peaks denoted by A, B, and C of ${\rm Re}(\sigma_{yy})$ 
originate from the direct inter-band transitions
$\phi^v_n \to \phi^c_{n}$ (A), $\phi^v_{n-1} \to \phi^c_{n-1}$ (B), and 
$\phi^v_{n+1} \to \phi^c_{n+1}$ (C), 
respectively, as shown in Fig.~\ref{fig:dc_arm}(b).
The energy position of the A peak corresponds to 
the energy gap ($E_{\rm gap}$) of 0.66 eV for 
an $N=15$ armchair nanoribbon.
Both a large density of states at the band edge ($k= 0$)
of each subband and the pseudospin are important 
as regards enhancing these peak intensities.

\begin{figure}[htbp]
 \begin{center}
  \includegraphics[scale=0.4]{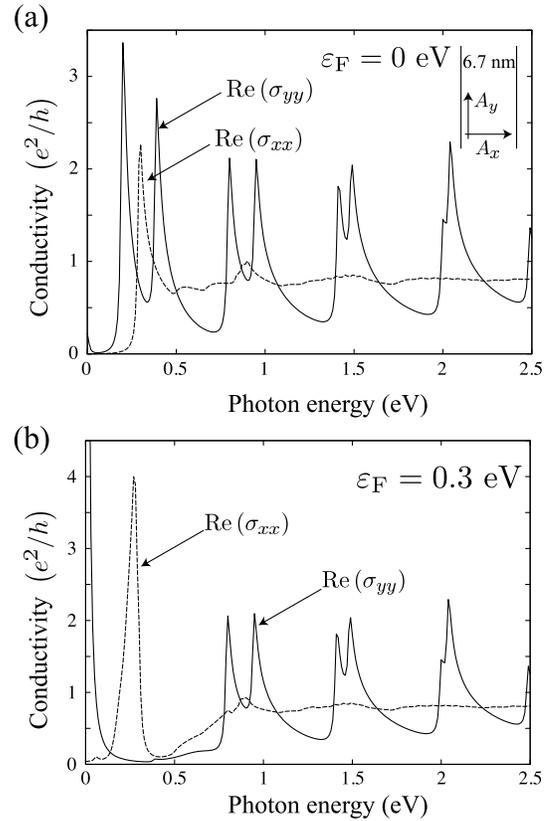}
 \end{center}
 \caption{${\rm Re}(\sigma_{\alpha\alpha})$ of an $N=55$ armchair
 nanoribbon with an energy gap about 0.2 eV, calculated for (a)
 $\varepsilon_{\rm F}=0$ eV and (b) 
 $\varepsilon_{\rm F}=0.3$ eV.
 }
 \label{fig:dc_arm_N55}
\end{figure}

The dashed curve in Fig.~\ref{fig:dc_arm}(a) 
represents ${\rm Re}(\sigma_{xx})$.
As we have seen in Eq.~(\ref{eq:interx}), 
the absorption of $x$-polarized light 
is the result of an indirect inter-band transition.
Thus, the peak denoted by D of ${\rm Re}(\sigma_{xx})$ 
is due to the transitions 
$\phi^v_n \to \phi^c_{n-1}$ and $\phi^v_{n-1} \to \phi^c_{n}$,
as shown in Fig.~\ref{fig:dc_arm}(c).
Because the resonance energy ($\varepsilon^c_n - \varepsilon^v_{n-1}$)
is between $\varepsilon^c_n - \varepsilon^v_n$ and $\varepsilon^c_{n-1} - \varepsilon^v_{n-1}$,
which are the resonance energies for the direct inter-band transitions A
and B, the peak D appears between the A and B peaks.
Note that another indirect transition $\phi^v_n \to \phi^c_{n+1}$ 
is also allowed by momentum conservation. 
However, the pseudospin strongly suppresses the backward transition for
the states at the band edge ($k=0$) and 
the large density of states does not result in an absorption peak.
Hence, there is no prominent peak for ${\rm Re}(\sigma_{xx})$ 
between the B and C peaks.
Moreover,
the peak corresponding to the lowest-energy backward transition is the most
prominent, and the peak for the second lowest (and higher) energy transition, 
such as $\phi^v_n \to \phi^c_{n-3}$ and $\phi^v_{n+1} \to \phi^c_{n-2}$,
is generally suppressed by the momentum conservation giving rise to
the suppression factor of $(m-n)^{-1}$ in Eq.~(\ref{eq:interx}).
To clarify this point,
we show the calculated ${\rm Re}(\sigma_{xx})$
for an $N=55$ armchair nanoribbon with $E_{\rm gap}=0.2$ eV 
in Fig.~\ref{fig:dc_arm_N55}(a).
For ${\rm Re}(\sigma_{xx})$,
there is only a peak on the low-energy side.

It is clear that 
the absorption spectrum of an armchair nanoribbon exhibits
strong anisotropy for the polarization
direction of incident light.~\cite{gundra11}
More specifically, a number of peaks appear when 
the polarization is parallel to the armchair nanoribbon, while
only a single prominent peak appears
on the low-energy side 
when the polarization is set
perpendicular to the ribbon.
Because the energy gap of (semiconducting) armchair nanoribbons
behaves as $E_{\rm gap} \propto N^{-1}$,~\cite{son06_energ,white07}
the energy position of the peak for perpendicular polarization
is inversely proportional to the width
of the armchair nanoribbon.

Since $\varepsilon_{\rm F}$ is generally not zero,
it is meaningful to point out that the absorption of low-energy photons
is sensitive to the position of $\varepsilon_{\rm F}$.
To explain this feature, we show the dynamical conductivity
for an $N=55$ armchair nanoribbon with $\varepsilon_{\rm F}=0.3$ eV 
in Fig.~\ref{fig:dc_arm_N55}(b).
For $E_{\rm gap} < \hbar \omega < 2|\varepsilon_{\rm F}|$,
${\rm Re}(\sigma_{yy})$ 
is suppressed by the Pauli exclusion principle, while a peak
remains for ${\rm Re}(\sigma_{xx})$.
This peak is attributed to forward intra-band scattering.
In addition, 
the Drude peak at $\hbar \omega = 0$ 
appears only for $y$-polarized light.
This feature arises from the fact that 
the diagonal matrix element of 
$v_y$ can take a non-zero value, while that of $v_x$ is zero,
as we have seen in 
Eqs.~(\ref{eq:ve1}) and (\ref{eq:ve}).

\section{Zigzag nanoribbon}\label{sec:zigrib}

In this section, 
we focus on zigzag nanoribbons.
We show that the optical properties of zigzag nanoribbons are
quantitatively different from those of armchair nanoribbons.
For example, when the polarization of an incident light is parallel to
the zigzag nanoribbons, the allowed inter-band transitions are indirect
transitions. 
On the other hand, when the light polarization is
perpendicular to the ribbons, a direct inter-band transition is allowed.
The behavior of zigzag nanoribbons contrasts with that of armchair
nanoribbons.

\subsection{Electron wavefunction}

\begin{figure}[htbp]
 \begin{center}
  \includegraphics[scale=0.45]{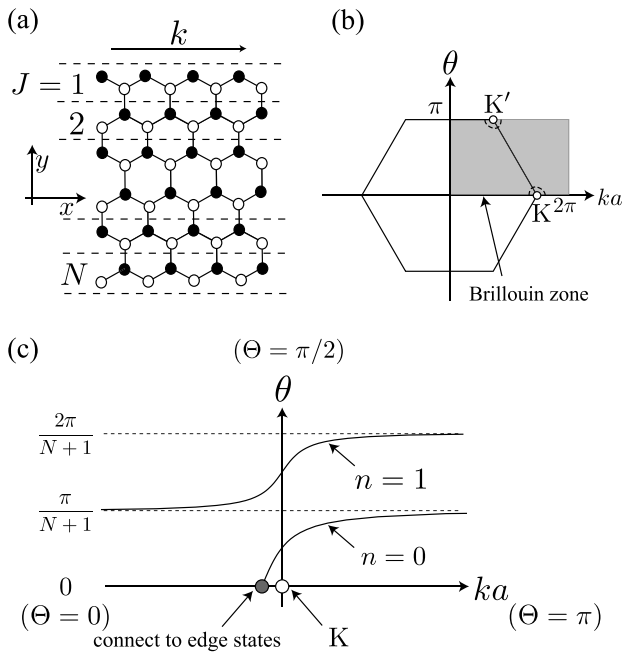}
 \end{center}
 \caption{ (a) The structure of an $N$ zigzag nanoribbon.
 (b) The BZ of a zigzag nanoribbon: 
 $\theta\in(0,\pi)$ and $ka\in[0,2\pi)$.
 The empty circles show the positions of the K and K$'$ points.
 The dashed semicircles around the K and K$'$ points represent 
 an energy contour (1 eV).
 (c) The plot of $\theta_n(k)$ for an $N$ zigzag nanoribbon. 
 }
 \label{fig:zig}
\end{figure}

The energy dispersion relation for zigzag nanoribbons
is given by 
\begin{align}
 \varepsilon^s(k,\theta) = 
 s\gamma_0 \sqrt{1 + 4 \cos^2\left(\frac{ka}{2}\right) + 4
 \cos\left(\frac{ka}{2}\right) \cos \theta},
 \label{eq:enez}
\end{align}
where $k$ is the wave vector along the zigzag edge, 
and $\theta$ denotes the phase
in the direction perpendicular to the edge 
[see Fig.~\ref{fig:zig}(a)].
By putting $(k,\theta) = (4\pi/3a,0)$ and $(2\pi/3a,\pi)$
into Eq.~(\ref{eq:enez}), we have $\varepsilon^s(k,\theta)=0$.
The two Dirac points are non-equivalent, i.e., 
the BZ of zigzag nanoribbons 
contains the two independent Dirac points (K and K$'$)
[see Fig.~\ref{fig:zig}(b)].
This contrasts with the fact that 
the BZ of armchair nanoribbons has a single Dirac point.
The difference in the number of Dirac points is because
the reflection of an electronic wave 
at the zigzag edge constitutes intravalley scattering, while the
reflection at the armchair edge is intervalley scattering.

The wave function is expressed by
\begin{align}
 \varphi^s_{J}(k,\theta) = \frac{1}{2\sqrt{N}} 
 \left\{
 e^{iJ\theta}
 \begin{pmatrix}
  e^{-i\Theta(k,\theta)} \cr s 
 \end{pmatrix}
 - e^{-iJ\theta}
 \begin{pmatrix}
  e^{i\Theta(k,\theta)} \cr s 
 \end{pmatrix}
 \right\},
\label{eq:wfzig}
\end{align}
where $J$ ($=1,\ldots,N$) is a coordinate 
perpendicular to the zigzag edge
[see Fig.~\ref{fig:zig}(a)].
A detailed derivation of the wave function including the pseudospin is
given in Ref.~\onlinecite{sasaki09}.
The wave function Eq.~(\ref{eq:wfzig}) is a standing wave 
formed by the superposition of two waves 
propagating in opposite directions.
The first term on the right-hand side of Eq.~(\ref{eq:wfzig})
represents an incident wave, and the second term 
corresponds to an edge reflected wave. 
The phase $\Theta(k,\theta)$ 
is defined through
$|\varepsilon^s(k,\theta)| e^{-i\Theta(k,\theta)} = 2
\cos\left(ka/2\right) + e^{-i\theta}$,~\cite{sasaki09}
by which we can see that
$\Theta(k,\theta)$ corresponds to the polar angle defined 
with respect to the K or K$'$ point.
For the K point, $\Theta(k,\theta)$ 
is given as shown in Fig.~\ref{fig:zig}(c).
Note that 
the signs in front of $\Theta(k,\theta)$ for the constituent waves are opposite.
This feature can be understood in terms of the reflection of pseudospin, namely, 
the zigzag edge alters the direction of the pseudospin
of an incident wave.~\cite{sasaki10-jpsj} 
By contrast, the armchair edge does not 
change the direction of the pseudospin.
In fact, in this case, we can reproduce 
the wave function for armchair edge Eq.~(\ref{eq:bw}) as
\begin{align}
 e^{iJ\theta}
 \begin{pmatrix}
  e^{-i\Theta} \cr s 
 \end{pmatrix}
 - e^{-iJ\theta}
 \begin{pmatrix}
  e^{-i\Theta} \cr s 
 \end{pmatrix}
 \propto 
 \sin(J\theta)
  \begin{pmatrix}
  e^{-i\Theta} \cr s 
 \end{pmatrix}.
 \label{eq:armwf-rec}
\end{align}
It is the characteristic feature of the zigzag edge
that the reflection of the momentum of the incident wave
$(\theta \to -\theta)$ accompanies the change of the pseudospin
$(\Theta \to -\Theta)$.
This correlation between momentum and pseudospin
does not exist for armchair nanoribbons.

The wave function has two components
\begin{align}
 \varphi^s_{J}=
 \begin{pmatrix}
  \varphi^s_{{\rm A},J} \cr
  \varphi^s_{{\rm B},J}
 \end{pmatrix}=
 \frac{i}{\sqrt{N}}
 \begin{pmatrix}
  \sin(J\theta-\Theta(k,\theta)) \cr s \sin (J\theta)
 \end{pmatrix},
\end{align}
where the first (second) component represents 
the amplitude at A-atom (B-atom) [see Fig.~\ref{fig:zig}(a)].
Note that the minus sign in front of the 
second term on the right-hand side of Eq.~(\ref{eq:wfzig})
ensures that the wave function satisfies the boundary condition 
at $J=0$ given by $\varphi^s_{{\rm B},J=0}=0$.
The phase $\theta$ is quantized as $\theta_n$
by the boundary condition at $J=N+1$, $\varphi^s_{{\rm A},J=N+1}=0$.
This boundary condition 
gives the constraint condition between $k$ and $\theta_n$,
$\Theta(k,\theta_n) - (N+1)\theta_n = -n\pi$, which is rewritten as
\begin{align}
 \theta_n = \frac{n\pi+\Theta(k,\theta_n)}{N+1}.
 \label{eq:Thetan}
\end{align}
We plot $\theta_n$ with $n=0,1$
as a function of $k$ in Fig.~\ref{fig:zig}(c), where 
$\theta_n$ is a curved rather than a straight line.~\cite{sasaki10-forward}
This feature also contrasts with that of an armchair nanoribbon,
for which the quantized $\theta_n$ is independent of $k$
[see Eq.~(\ref{eq:theta})].
Note that the curved line of $\theta_n(k)$ with $n=0$
reaches the $k$-axis near the Dirac point. 
It can be shown that $\theta_0(k)$ 
acquires an imaginary part after crossing the $k$-axis
as $\theta_0(k)=i/\xi(k)$.
In Eq.~(\ref{eq:wfzig}),
$\xi(k)$ corresponds to the localization length.
By putting $\theta_0(k)=i/\xi(k)$ into Eq.~(\ref{eq:Thetan}), 
we see that $\Theta_0(k,\theta_0(k))$ also acquires an imaginary part
as $\Theta_0(k,\theta_0(k))=i(N+1)/\xi(k)$.
This means that the pseudospin is polarized by the localization.
The localized states near the zigzag edge 
are known as the edge states.~\cite{tanaka87,fujita96} 
The curved line is essential to the existence of the edge
localized states in the zigzag nanoribbons.

\subsection{Selection rule}\label{sec:optzig}

We start by showing 
the velocity matrix elements 
with respect to the inter-band transition 
(see Appendix~\ref{app:zig} for derivation),
\begin{align}
\begin{split}
 & \langle \varphi^{c}_m | v_x  | \varphi^{v}_n \rangle 
 = i v_{\rm F}
 \langle \varphi^{c}_m | \sigma_y  | \varphi^{c}_n \rangle, \\
 & \langle \varphi^{c}_m | v_y  | \varphi^{v}_n \rangle 
 = i v_{\rm F}
 \langle \varphi^{c}_m | \sigma_x | \varphi^{c}_n \rangle.
\end{split}
 \label{eq:vqzig}
\end{align}
The calculation for the pseudospin of the standing wave,
on the right-hand side of Eq.~(\ref{eq:vqzig}),
is straightforward with the use of Eqs.~(\ref{eq:wfzig})
and (\ref{eq:Thetan}).
The results are
\begin{widetext}
\begin{align}
 \langle \varphi^{c}_m | \sigma_x | \varphi^{c}_n \rangle 
 =&-\sin\Theta_m \left\{ \frac{\sin^2 \left[
 \frac{(n+m)\pi+(\Theta_n+\Theta_m)}{2}\right]
 }{(n+m)\pi+(\Theta_n+\Theta_m)} + 
 \frac{\sin^2 \left[
 \frac{(n-m)\pi+(\Theta_n-\Theta_m)}{2} \right]
 }{(n-m)\pi+(\Theta_n-\Theta_m)}
 \right\} \nn \\
 &- \frac{\cos\Theta_m}{2} \left\{ \frac{\sin\left[
 (n+m)\pi+(\Theta_n+\Theta_m)\right]
 }{(n+m)\pi+(\Theta_n+\Theta_m)} - 
 \frac{\sin \left[
 (n-m)\pi+(\Theta_n-\Theta_m) \right]
 }{(n-m)\pi+(\Theta_n-\Theta_m)}
 \right\} + (n\leftrightarrow m),
 \label{eq:sigmatrix} 
\end{align}
\begin{align}
 i\langle \varphi^{c}_m | \sigma_y | \varphi^{c}_n \rangle 
 =&
 -\sin\Theta_m \left\{ \frac{\sin^2 \left[
 \frac{(n+m)\pi+(\Theta_n+\Theta_m)}{2}\right]
 }{(n+m)\pi+(\Theta_n+\Theta_m)} + 
 \frac{\sin^2 \left[
 \frac{(n-m)\pi+(\Theta_n-\Theta_m)}{2} \right]
 }{(n-m)\pi+(\Theta_n-\Theta_m)}
 \right\} \nn \\ 
 &- \frac{\cos\Theta_m}{2} \left\{ \frac{\sin\left[
 (n+m)\pi+(\Theta_n+\Theta_m)\right]
 }{(n+m)\pi+(\Theta_n+\Theta_m)} - 
 \frac{\sin \left[
 (n-m)\pi+(\Theta_n-\Theta_m) \right]
 }{(n-m)\pi+(\Theta_n-\Theta_m)}
 \right\} - (n\leftrightarrow m).
 \label{eq:sigmatriy}
\end{align}
\end{widetext}

%

First, we consider a case where
the polarization of the incident light 
is parallel to the zigzag nanoribbon.
By setting $m=n$ in Eq.~(\ref{eq:sigmatriy}), 
we have
$\langle \varphi^{c}_n | \sigma_y | \varphi^{c}_n \rangle =0$.
Since $\langle \varphi^{c}_n | \sigma_y | \varphi^{c}_n \rangle=0$ leads to
$\langle \varphi^{c}_n | v_x | \varphi^{v}_n \rangle=0$ 
in Eq.~(\ref{eq:vqzig}), 
we conclude that the $x$-polarized light 
does not cause a direct inter-band transition.
Therefore, the possible inter-band transition is an indirect one.
To further explore the inter-band transitions 
produced by the $x$-polarized light,
let us examine the electrons with $\Theta= \pi/2$ (or $k= 4\pi/3a$).
By putting
$\Theta_n = \pi/2$ and $\Theta_m = \pi/2$
into Eq.~(\ref{eq:sigmatriy}),
we obtain
\begin{align}
 \langle \varphi^{c}_m | v_x | \varphi^{v}_n \rangle =
 v_{\rm F} 
 \frac{2}{\pi}\frac{\sin^2 \left(\frac{\pi}{2}\Delta n \right)}{\Delta n}.
 \label{eq:sig_zxy}
\end{align}
Equation~(\ref{eq:sig_zxy}) suggests that 
the $x$-polarized light induces indirect transitions
when $m-n$ $(\equiv \Delta n)$ is an odd number.
The inter-band transitions with $\Delta n=\pm 1$
have advantage over the transitions with $\Delta n=\pm 3, 5, \ldots$ 
in producing prominent peaks in the dynamical conductivity since the
suppression due to the momentum conservation is minimum 
when $\Delta n=\pm 1$.

Next, we consider a case where
the polarization of the incident light 
is perpendicular to the zigzag nanoribbon.
When $m=n$, Eq.~(\ref{eq:sigmatrix})
leads to 
\begin{align}
 \langle \varphi^{c}_n | \sigma_x | \varphi^{c}_n \rangle 
 =\cos\Theta_n -\frac{\sin\Theta_n}{n\pi+\Theta_n}.
\end{align}
The second term on the right-hand side is suppressed by 
the factor of $(\pi n)^{-1}$ when $n \ge 1$.
For $n=0$, 
$\langle \varphi^{c}_n | \sigma_x | \varphi^{c}_n \rangle$ 
is zero when $\Theta_0=0$ and $-1$ when $\Theta_0=\pi$.
Thus, the $y$-polarized light gives rise to direct inter-band
transitions, by which 
the electrons near the $k$-axis ($\Theta=0$ or $\pi$)
are selectively excited.
Since there is not a large density of states for the states near the
$k$-axis, the direct inter-band transition does not result in a
prominent absorption peak.

\begin{figure}[htbp]
 \begin{center}
  \includegraphics[scale=0.4]{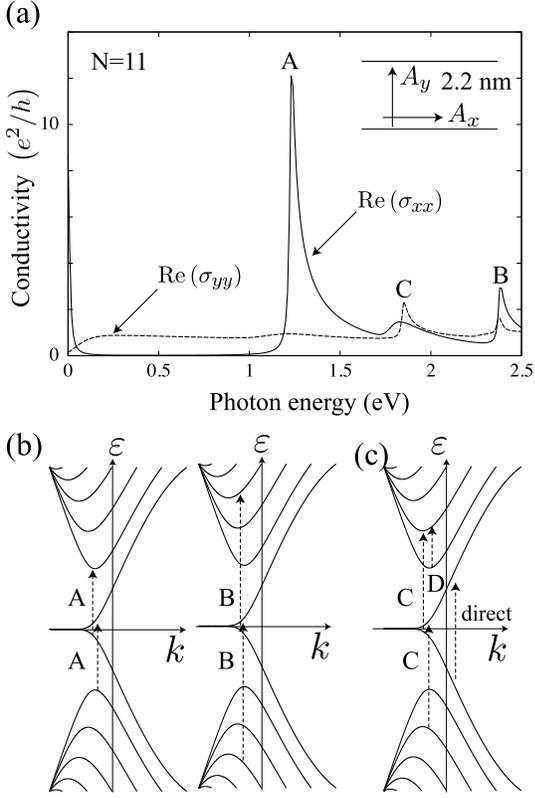}
 \end{center}
 \caption{ (a) Calculated ${\rm Re}(\sigma_{\alpha\alpha})$
 for an $N=11$ zigzag nanoribbon.
 The optical transition processes 
 relevant to (b) ${\rm Re}(\sigma_{xx})$ and 
 (c) ${\rm Re}(\sigma_{yy})$.
 The flat band at the band center represents the energy dispersion of 
 the edge states. 
 }
 \label{fig:dc_zig}
\end{figure}

By putting
$\Theta_n = \pi/2$ and $\Theta_m = \pi/2$
into Eq.~(\ref{eq:sigmatrix}),
we obtain
\begin{align}
 \langle \varphi^{c}_m | v_y | \varphi^{v}_n \rangle =
 -iv_{\rm F} \frac{2}{\pi} \frac{\sin^2\left[\frac{\pi}{2}(n+m+1)\right]}{(n+m+1)}
 \label{eq:sig_y}
\end{align}
Equation~(\ref{eq:sig_y}) suggests that 
the $y$-polarized light gives rise to an indirect inter-band transition 
when $n + m$ is an even number.
For example, the transition $\varphi^v_0 \to \varphi^c_2$
[which corresponds to 
the peak C in Fig.~\ref{fig:dc_zig}(c)]
satisfies this condition.
However, the amplitude of this process is small due to the
suppression by momentum conservation.
The indirect inter-band transitions caused by $y$-polarized
light can be neglected.

We have shown that 
the $x$-polarized light (parallel to the zigzag edge)
results in indirect inter-band transitions ($\Delta n=\pm 1$)
for the states near the band edges.~\cite{hsu07}
The $y$-polarized light induces direct inter-band transitions
($\Delta n=0$) selectively for the states near the $k$-axis.
The polarization dependence of the inter-band optical transition
in zigzag nanoribbons exhibits a 90$^\circ$ phase shift 
with respect to that derived for armchair nanoribbons.

The velocity matrix elements 
for the intra-band transition are written as
\begin{align}
\begin{split}
 & \langle \varphi^{c}_m | v_x | \varphi^{c}_n \rangle 
 = -v_{\rm F}
 \langle \varphi^{c}_m |\sigma_x | \varphi^{c}_n \rangle,
 \\
 & \langle \varphi^{c}_m | v_y  | \varphi^{c}_n \rangle 
 = v_{\rm F}
 \langle \varphi^{c}_m |\sigma_y | \varphi^{c}_n \rangle.
\end{split}
 \label{eq:vezig}
\end{align}
From Eqs.~(\ref{eq:vezig}) and (\ref{eq:vqzig}),
we see that 
\begin{align}
\begin{split}
 & |\langle \varphi^{c}_m | v_x  | \varphi^{v}_n \rangle |
 = |\langle \varphi^{c}_m | v_y  | \varphi^{c}_n \rangle |, \\
 & |\langle \varphi^{c}_m | v_y  | \varphi^{v}_n \rangle |
 =|\langle \varphi^{c}_m | v_x  | \varphi^{c}_n \rangle |.
\end{split}
\end{align}
The equations mean that
the selection rule for intra-band transitions can be obtained from 
that for inter-band transitions by changing the polarization direction.
For the intra-band transitions,
the $x$-polarized light results in 
a direct transition ($\Delta n=0$), while 
the $y$-polarized light results in 
an indirect transition ($\Delta n=\pm 1$).

\subsection{Dynamical conductivity}\label{ssec:dczig}

In Fig.~\ref{fig:dc_zig}(a), 
we show the calculated ${\rm Re}(\sigma_{\alpha\alpha})$
for an $N=11$ zigzag nanoribbon. 
The width of the ribbon is 2.2 nm.
The solid curve represents 
${\rm Re}(\sigma_{xx})$, where 
the A and B peaks 
are attributed to the transitions
shown in Fig.~\ref{fig:dc_zig}(b).
The change in the wavenumber, 
which is relevant to the A peak,
is $\Delta n =\pm 1$, while that relevant to the B peak is $\Delta n =\pm 3$.
The intensity of the B peak is much smaller than that of the A peak
due to suppression by momentum conservation.
The dashed curve represents ${\rm Re}(\sigma_{yy})$, where 
the small C peak is due to the transitions shown in Fig.~\ref{fig:dc_zig}(c).
The finite conductivity at low energy below the C peak
originates from the direct transition shown in Fig.~\ref{fig:dc_zig}(c).
The conductivity is suppressed near $\hbar \omega=0$.
This feature can be explained by the pseudospin of the edge states.
Later we discuss the fact that 
the velocity matrix element between two edge states vanishes
because the edge states are pseudospin polarized states
(i.e., eigen state of $\sigma_z$), so that $\langle \sigma_x \rangle_{nn}=0$.

\begin{figure}[htbp]
 \begin{center}
  \includegraphics[scale=0.4]{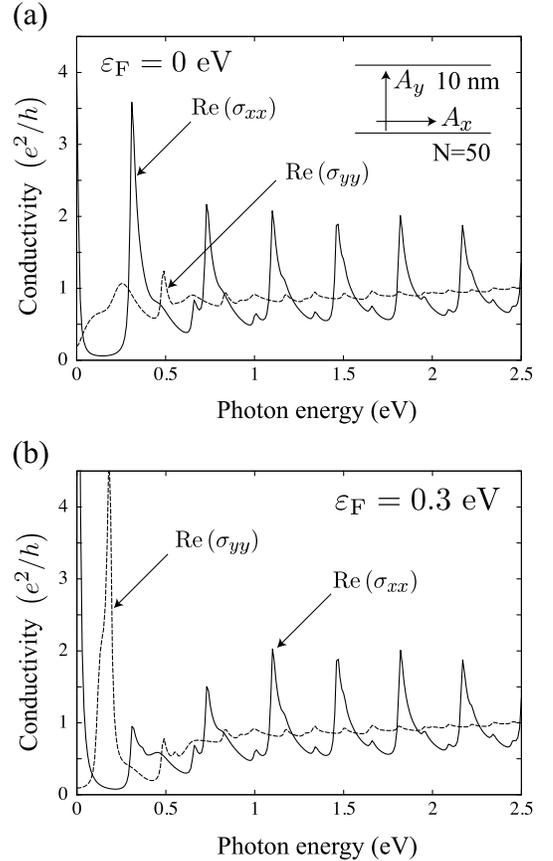}
 \end{center}
 \caption{The calculated ${\rm Re}(\sigma_{\alpha\alpha})$
 of an $N=50$ zigzag nanoribbon for different Fermi energies:
 (a) $\varepsilon_{\rm F}=0$ eV and (b) $\varepsilon_{\rm F}=0.3$ eV. 
 }
 \label{fig:dc_zig_N50}
\end{figure}

Note that the indirect inter-band transitions 
A, B, and C in Fig.~\ref{fig:dc_zig}(b) and (c),
involve the states near the Dirac point.
Hence, we can expect the dynamical conductivity to be
sensitive to the position of $\varepsilon_{\rm F}$.
To elucidate this point,
we show the calculated ${\rm Re}(\sigma_{\alpha\alpha})$
for an $N=50$ zigzag nanoribbon with different 
$\varepsilon_{\rm F}$
in Fig.~\ref{fig:dc_zig_N50}(a) [$\varepsilon_{\rm F}=0$ eV]
and (b) [$\varepsilon_{\rm F}=0.3$ eV].
In Fig.~\ref{fig:dc_zig_N50}(a),
the most prominent peak appears for ${\rm Re}(\sigma_{xx})$, which 
corresponds to the transition A ($\varphi^v_0 \to \varphi^c_{1}$)
shown in Fig.~\ref{fig:dc_zig}(b). 
The other peaks are due to 
$\varphi^v_n \to \varphi^c_{n+1}$ with $n=1,2,\ldots$.
By contrast, at a low energy in Fig.~\ref{fig:dc_zig_N50}(b),
the peak intensity is suppressed for ${\rm Re}(\sigma_{xx})$
and the most prominent peak appears for ${\rm Re}(\sigma_{yy})$.
This change of the polarization dependence 
is because intra-band transitions, 
such as D shown in Fig.~\ref{fig:dc_zig}(c), are activated
by $y$-polarized light.
Finally, the Drude peak appears only when the polarization is set parallel to
the zigzag nanoribbon, because the $x$-polarized light provides a direct
intra-band transition.
This polarization dependence of the Drude peak 
is analogous to the case of armchair nanoribbon.

\section{Irregular edge}\label{sec:mixed}

In this section, 
we study the effect of edge irregularities
on the dynamical conductivity 
by employing a numerical simulation based on a tight-binding model.
Although the variety of irregular edge structures that we consider
is quite limited, the results help us 
to understand the way in which 
the absorption spectra are changed by irregularities.

\begin{figure}[htbp]
 \begin{center}
  \includegraphics[scale=0.35]{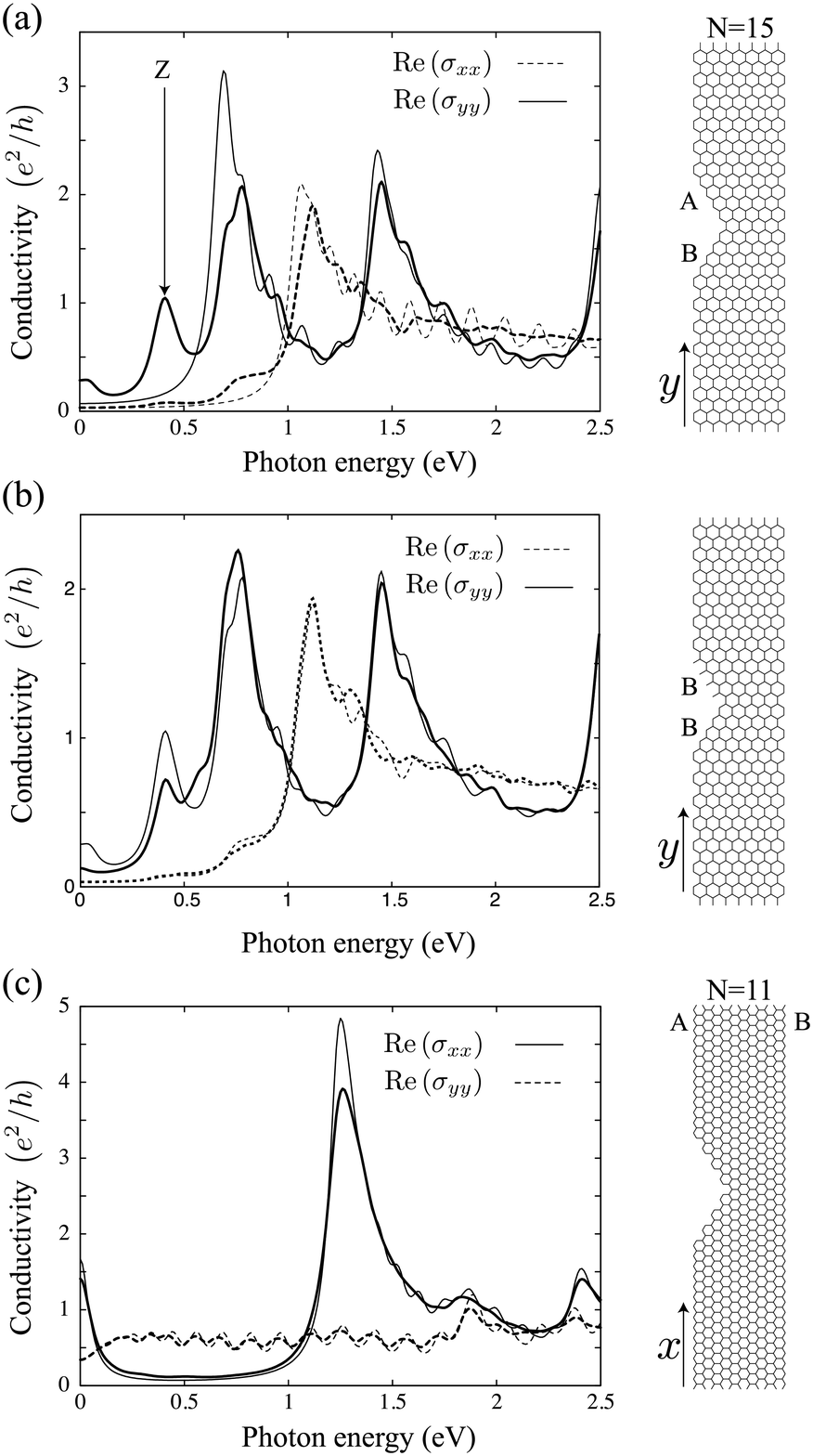}
 \end{center}
 \caption{The calculated ${\rm Re}(\sigma_{\alpha\alpha})$ 
 in the presence (thick lines) or absence (thin lines) of irregularities for 
 (a) $N=15$ armchair and (c) $N=11$ zigzag nanoribbons.
 (b) compares ${\rm Re}(\sigma_{\alpha\alpha})$ 
 in the presence of an irregularity consisting of two sublattices 
 (thin lines) with that consisting of one sublattice (thick lines).
 In the simulation, we assume that 
 the length of the nanoribbon is 40 nm, $\delta=50$ meV,
 and $\varepsilon_{\rm F}=0$ eV.
 }
 \label{fig:dc_mix}
\end{figure}

Figure~\ref{fig:dc_mix}(a) displays
the calculated ${\rm Re}(\sigma_{\alpha\alpha})$
for a defective $N=15$ armchair nanoribbon (thick lines).
We introduced irregularities consisting of zigzag edges
along the armchair edge of the ribbon.
One of the zigzag parts is shown in the right-hand side of
Fig.~\ref{fig:dc_mix}(a).
The size of the zigzag part is about $\Delta x=5$ \AA \
and $\Delta y=2$ nm.
For comparison,
we show ${\rm Re}(\sigma_{\alpha\alpha})$
for a pure armchair nanoribbon without any irregularities
(thin lines) in Fig.~\ref{fig:dc_mix}(a).
These results show that 
the irregularities do not induce any significant change
in the ${\rm Re}(\sigma_{xx})$
of the regular armchair nanoribbon.
A notable effect of the irregularities 
appears for ${\rm Re}(\sigma_{yy})$
as the new peak at 0.4 eV, denoted by Z.
Since the original pure armchair nanoribbon 
has an energy gap of 0.66 eV, 
the Z peak originates from irregularities 
consisting of the zigzag parts. 
Note that 
the appearance of the Z peak 
accompanies a reduction in the peak
intensity at 0.66 eV.

Because
a small segment of the zigzag edge about 1 nm in length 
creates a zero-energy edge state in the energy spectrum,~\cite{nakada96}
it is reasonable to consider that 
the Z peak is relevant to the edge state caused by the
irregularity.
Moreover, the reduction in the peak intensity at 0.66 eV suggests that 
the optical transition from the edge state to the state at the band edge of the
first subband is activated.
This consideration is consistent with the fact that
the Z peak position (0.4 eV) is approximately half of the energy 
band gap of the pure armchair nanoribbon (0.66/2 eV).

An interesting point here is that 
the edge state existing near the zigzag irregularity
can contribute to the optical transition, 
on the other hand, the edge state
in pure zigzag nanoribbons does not.
That is, the edge state near a zigzag irregularity is
not identical to the edge state in a pure zigzag nanoribbon.
For a pure zigzag nanoribbon,
the edge atoms at $J=0$ are all A-atoms, while 
those at $J=N$ are all B-atoms [see Fig.~\ref{fig:zig}(a)].
As a result, the wave function of the edge state localized at $J=0$ 
is written as
\begin{align}
 \varphi_{\rm A} \propto e^{-\frac{J}{\xi}}
 \begin{pmatrix}
  1 \cr 0
 \end{pmatrix},
\end{align}
while the wave function of the edge state localized at $J=N$
is written as
\begin{align}
 \varphi_{\rm B} \propto e^{-\frac{N-J}{\xi}}
 \begin{pmatrix}
  0 \cr 1
 \end{pmatrix},
\end{align}
where $\xi$ is the localization length.
Note that these edge states
have an amplitude only on the A-atoms or B-atoms 
depending on the edge atom.
Since the electron-light matrix element is proportional to the matrix
element of $\sigma_x$ or $\sigma_y$, the matrix elements with respect to
the pseudospin polarized edge states vanish as
\begin{align}
 \begin{pmatrix}
  1 & 0
 \end{pmatrix}
 \sigma_i
 \begin{pmatrix}
  1 \cr 0
 \end{pmatrix}
 =0  \ \ (i=x,y).
\end{align}
One might consider that the matrix element of $\sigma_i$
between the edge state localized at $J=0$ and that localized at $J=N$
can be non-zero because
\begin{align}
 \begin{pmatrix}
  0 & 1
 \end{pmatrix}
 \sigma_i
 \begin{pmatrix}
  1 \cr 0
 \end{pmatrix}
 \ne 0.
\end{align}
However, such matrix elements are also strongly suppressed
because $\xi < N$ is satisfied and 
there is little spatial overlap between the two edge states.
Therefore, the edge state in a pure zigzag nanoribbon
does not contribute to the optical transition.

By contrast, the edge atoms of the zigzag irregularity 
include equal numbers of A-atoms and B-atoms, 
and the A- and B-edge atoms are close in space. 
Thus, the edge state at the A-edge atoms 
[$\varphi_{\rm A}({\bf r})$]
can interfere with the other
edge state at the B-atoms 
[$\varphi_{\rm B}({\bf r})$]
to form a new edge state pair
[$\varphi^{c,v}({\bf r})=\varphi_{\rm A}({\bf r})\pm\varphi_{\rm B}({\bf r})$].
This feature of the interference between two sublattices
is essential to the optical transition.
This consideration is consistent with the result 
shown in Fig.~\ref{fig:dc_mix}(b), 
where the intensity of the Z peak decreases when
the zigzag edge part consists only of one sublattice.~\cite{klein94}
Note also that the inter-band transition $\varphi^v({\bf r}) \to
\varphi^c({\bf r})$ for the edge states contributes 
to the absorption spectrum at zero-energy ($\hbar \omega\approx 0$)
as shown in Fig.~\ref{fig:dc_mix}(a).

In Fig.~\ref{fig:dc_mix}(c),
we show ${\rm Re}(\sigma_{\alpha\alpha})$
for a defective $N=11$ zigzag nanoribbon (thick lines)
and that for a pure one (thin lines). 
The irregularities consist of armchair edges.
One of the armchair parts is shown in the right-hand side of
Fig.~\ref{fig:dc_mix}(c).
The irregularities do not result in a notable change of the absorption spectrum.
This feature can be rationalized 
by using the fact that the armchair edge
does not change the pseudospin
via the reflection of the incident electron wave
[see Eq.~(\ref{eq:armwf-rec})].

The numerical results presented in this section 
suggest that for a defective armchair nanoribbon,
the absorption spectrum of transversely ($x$) polarized light 
is robust against the presence of irregularities, while
the absorption spectrum of low-energy longitudinally ($y$) polarized
light is sensitive to them.
For a defective zigzag nanoribbon,
the absorption spectrum appears insensitive to the irregularities.

\section{Discussion}\label{sec:dis}

The optical selection rule of armchair nanoribbons
reminds us of the optical selection rule of single-wall
carbon nanotubes (SWNTs).~\cite{ajiki94}
The selection rule of SWNTs shows that 
light polarized parallel (perpendicular) to the axis 
results in a direct (indirect) inter-band transition 
satisfying $\Delta n=0$ ($\Delta n=\pm 1$).
It is known that the indirect transition
originates from the cylindrical topology of SWNTs.
Although the indirect transition is interesting, 
the corresponding absorption peak 
is suppressed by the depolarization effect.~\cite{ajiki94}
Indeed, the experimental results 
for SWNTs do not show absorption peaks 
relevant to the indirect inter-band transition.
This is consistent with the presence of a depolarization
field.~\cite{hwang00,ichida01,ichida04}
While the topology of nanoribbons
is quite different from that of SWNTs, 
there is a possibility that a depolarization
field suppresses the absorption of the $x$-polarized light
for armchair nanoribbons.
The selection rule for zigzag nanoribbons 
is distinct from that for armchair nanoribbons, and 
the zigzag nanoribbons seem to be anomalous 
with respect to the depolarization effect.
The depolarization field
points along the $y$-axis, 
which is orthogonal to the direction of the polarization of the incident light.


The optical transition between two subbands $n$ and $m$ 
depends on changes in the wavenumber $\Delta n$ ($= m-n$)
and the pseudospin $\langle \sigma_i \rangle_{mn}$.
These factors $\Delta n$ and $\langle \sigma_i \rangle_{mn}$
could be treated independently 
in a discussion of the optical transitions
for armchair nanoribbons. On the other hand, 
the factors mix and cannot be treated
independently for zigzag nanoribbons.
The pseudospin can be changed by interactions 
that are not included in our analysis. 
For example, a potential difference between the A and B sublattices
is induced by the stack in the Bernal configuration and gives rise
to a polarization of pseudospin for the states near the Dirac
point.~\cite{sasaki11-armwf} 
Furthermore, the pseudospin can be related to real spin 
because the edge states in zigzag nanoribbons 
are prone to be spin polarized if we take into account the Coulomb
interaction.~\cite{fujita96,son06_energ}
Although the selection rules that we have obtained 
provide the basis for exploring the optical
phenomena exhibited by graphene nanoribbons,
it is possible that these interactions 
modify the absorption peaks present in this paper.~\cite{gundra11}

\section{conclusion}\label{sec:con}


We have paid considerable attention to 
the indirect optical transitions satisfying $\Delta n=\pm 1$.
An inter-band transition with $\Delta n=\pm 1$
is induced when 
the polarization of the incident light is perpendicular (parallel)
to the armchair (zigzag) edge, while 
an intra-band indirect transition is induced when 
the light polarization is perpendicular
to both the armchair and zigzag edges.
The indirect optical transitions can be traced to
the inter-edge electronic coherence. 
The signal of the indirect transition appears as a low-energy peak 
in absorption spectra.
The corresponding absorption peak remains 
in the presence of the irregularity and 
with different Fermi energy positions, while 
it may be suppressed by a depolarization field.

\section*{Acknowledgments}

K.S. is supported by 
a Grant-in-Aid for Specially Promoted Research
(Grant No.~23310083) from the Ministry of Education, Culture, Sports, Science
and Technology.

\appendix

\section{Derivations of Eqs.~(\ref{eq:ve1}), (\ref{eq:ve}), (\ref{eq:interx}),
 and (\ref{eq:intery})}\label{app:v}

In this appendix, 
we give the derivations of 
the matrix elements of the velocity operator $\bv$
for armchair nanoribbons. 
The operator $\bv$ is defined as
the first derivative of the nearest-neighbor tight-binding Hamiltonian, 
$H({\bf A})$, with respect to a vector potential, ${\bf A}$, as 
$-e \bv = \partial H({\bf A})/\partial {\bf A}|_{{\bf A}=0}$.
The electron-light interaction is given by
$H_{\rm em}=-e \bv\cdot {\bf A}$, and 
the matrix elements are written as
\begin{align}
 & \langle \phi^{c}_m | \left( -e \bv\cdot {\bf A} \right) | \phi^{s}_n \rangle 
 = -ie
 \frac{\gamma_0}{\hbar} \sum_{J=1}^N 
 \left[ \phi^c_{m,J} \right]^\dagger \times \nn \\
 & \left(
 \delta h_{J}^+ G^+ \phi^s_{n,J+1} + 
 \delta h_{J} \phi^s_{n,J} +
 \delta h_{J}^- G^- \phi^s_{n,J-1} \right).
 \label{eq:Hem}
\end{align}
On the right-hand side of Eq.~(\ref{eq:Hem}), 
we have defined the translational operators,
\begin{align}
 G^+ \equiv 
 \begin{pmatrix}
  e^{ikb} & 0 \cr 0 & 1
 \end{pmatrix},
 \ \
 G^- \equiv
 \begin{pmatrix}
  1 & 0 \cr 0 & e^{-ikb}
 \end{pmatrix},
\label{app:defG}
\end{align} 
and the 2$\times$2 matrices of 
the potential induced by ${\bf A}$ as
\begin{align}
\begin{split}
 & \delta h_{J} \equiv 
 \sigma_- {\bf A} \cdot {\bf R}_1
 - \sigma_+ {\bf A} \cdot {\bf R}_1,
 \\
 & \delta h_{J}^+ \equiv 
 \sigma_- {\bf A} \cdot {\bf R}_2
 - \sigma_+ {\bf A} \cdot {\bf R}_3,
 \\
 & \delta h_{J}^- \equiv 
 \sigma_- {\bf A} \cdot {\bf R}_3
 - \sigma_+ {\bf A} \cdot {\bf R}_2,
\end{split}
\label{app:hjA}
\end{align}
where
\begin{align}
 \sigma_+ =
 \begin{pmatrix}
  0 & 1 \cr 
  0 & 0
 \end{pmatrix},
 \ \ 
 \sigma_- =
 \begin{pmatrix}
  0 & 0 \cr 
  1 & 0
 \end{pmatrix}.
\label{app:spm}
\end{align}
In Eq.~(\ref{eq:Hem}),
the translational operators $G^\pm$ should be inserted 
between $\delta h_{J}^{\pm}$ and $\phi^s_{n,J\pm 1}$
because $\phi^s_{n,J}$ of Eq.~(\ref{eq:bw}) is defined along
the diagonal line of the boxes shown in Fig.~\ref{fig:arm}(a).
Although $G^\pm \approx \sigma_0$
for the states near the Dirac point,
we do not use such an approximation here and
keep the derivation mathematically exact.
In Eq.~(\ref{app:hjA}), 
${\bf R}_a$ ($a=1,2,3$) are the vectors pointing from an A-atom
to the nearest-neighbor B-atoms defined as follows:
\begin{align}
\begin{split}
 & {\bf R}_1=a_{\rm cc}{\bf e}_y, \\
 & {\bf R}_2=- \frac{\sqrt{3}}{2} a_{\rm cc}{\bf e}_x - \frac{1}{2} 
 a_{\rm cc}{\bf e}_y, \\
 & {\bf R}_3=\frac{\sqrt{3}}{2} a_{\rm cc}{\bf e}_x - \frac{1}{2}
 a_{\rm cc}{\bf e}_y, 
\end{split}
\label{app:Ra}
\end{align}
where ${\bf e}_x$ (${\bf e}_y$) 
is the dimensionless unit vector for the $x$-axis
($y$-axis) and $a_{\rm cc}$ is the bond length between nearest-neighbor
carbon atoms.
By putting Eq.~(\ref{app:Ra}) into Eq.~(\ref{app:hjA}), 
we have
\begin{align}
\begin{split}
 & \delta h_{J} = -i \sigma_y (A_y a_{\rm cc}),
 \\
 & \delta h_{J}^+ 
 = -\frac{\sqrt{3}}{2} \sigma_x  (A_x a_{\rm cc}) + \frac{i}{2} \sigma_y
 (A_y a_{\rm cc}),
 \\
 & \delta h_{J}^- 
 = \frac{\sqrt{3}}{2}  \sigma_x (A_x a_{\rm cc}) + \frac{i}{2} \sigma_y 
 (A_y a_{\rm cc}).
\end{split}
\label{app:hjarm}
\end{align}
It is possible to consider a case where
${\bf A}$ is spatially modulated.~\cite{gupta09} 
But, here we assume a spatially uniform vector potential ${\bf A}$, 
so that the potential does not depend on $J$.

By putting Eqs.~(\ref{app:defG}) and (\ref{app:hjarm})
into Eq.~(\ref{eq:Hem}), and by 
comparing the coefficient of $A_x$ ($A_y$) 
on both sides of the equation, 
we obtain
\begin{align}
\begin{split}
 \langle \phi^c_m| v_x | \phi^s_n \rangle
 &= v_{\rm F}
 \sum_{J=1}^N \left[ \phi^c_{m,J} \right]^\dagger 
 \frac{-i}{\sqrt{3}}\left\{
 T \phi^s_{n,J+1} - T^{-1} \phi^s_{n,J-1}
 \right\},
 \\
 \langle \phi^c_m| v_y | \phi^s_n \rangle
 &= v_{\rm F}
 \sum_{J=1}^N \left[ \phi^c_{m,J} \right]^\dagger 
 \left\{
 \sigma_y - i\frac{\varepsilon_n^s}{3} \sigma_z
 \right\}\phi^s_{n,J},
\end{split}
\label{eq:ve-1}
\end{align}
where 
the matrix $T$ ($T^{-1}$) is defined by 
$T \equiv \sigma_x G^+$ ($T^{-1} \equiv \sigma_x G^-$):
\begin{align}
 T =
 \begin{pmatrix}
  0 & 1 \cr e^{ikb} & 0
 \end{pmatrix}, \ \ 
 T^{-1} =
 \begin{pmatrix}
  0 & e^{-ikb} \cr 1 & 0
 \end{pmatrix}.
 \label{app:ttinv}
\end{align}
Note that 
the eigen equation 
for the electron in an armchair nanoribbon
is written in terms of these $T$, $T^{-1}$, 
and $K^s_n \equiv \sigma_x + \varepsilon^s_n \sigma_0$ as~\cite{sasaki11-armwf}
\begin{align}
 T \phi^s_{n,J+1} + K^s_n \phi^s_{n,J} + T^{-1} \phi^s_{n,J-1} = 0.
\end{align}
The energy eigen equation has been used 
to obtain $\langle \phi^c_m| v_y | \phi^s_n \rangle$ 
of Eq.~(\ref{eq:ve-1}).

Below, we execute the summation over $J$ 
in Eq.~(\ref{eq:ve-1}).
First, we consider
$\langle \phi^c_m| v_y | \phi^s_n \rangle$ of Eq.~(\ref{eq:ve-1}).
By putting Eq.~(\ref{eq:bw}) into Eq.~(\ref{eq:ve-1}), 
and by using the formula,
\begin{align}
 \frac{2}{N}\sum_{J=1}^N \sin (J\theta_m) \sin(J\theta_n)
 = \delta_{mn},
 \label{app:inphase}
\end{align}
we have
\begin{align}
\begin{split}
 & \langle \phi^c_m| v_y | \phi^c_n \rangle
 = v_{\rm F}\delta_{mn} \langle \sigma_y \rangle_{mn}, \\
 & \langle \phi^c_m| v_y | \phi^v_n \rangle
 = i v_{\rm F} \delta_{mn}
 \left\{ \langle \sigma_x \rangle_{mn} - \frac{\varepsilon_n^v}{3}
 \right\}.
\end{split}
\end{align}
These results are shown in Eqs.~(\ref{eq:ve}) and (\ref{eq:intery}).
In Eq.~(\ref{eq:intery}), 
the small term, $\varepsilon^v_n/3$, has been omitted. 
Next, we take $\langle \phi^c_m| v_x | \phi^s_n \rangle$
of Eq.~(\ref{eq:ve-1}).
By using Eq.~(\ref{eq:bw}), we obtain the equation,
\begin{align}
 &T \phi^s_{J+1} -T^{-1} \phi^s_{J-1} = \nn \\
 & 2 \sin\theta Te^{-ikl} \frac{1}{\sqrt{N}}e^{-ikl(J-1)} \cos(J\theta) 
 \begin{pmatrix}
  e^{-i\Theta} \cr s 
 \end{pmatrix}.
\label{app:Tphi}
\end{align}
By putting this into Eq.~(\ref{eq:ve-1}), we obtain
\begin{align}
 \langle \phi^c_m| v_x | \phi^s_n \rangle
 &=-iv_{\rm F}  \frac{2\sin
 \theta_n}{\sqrt{3}}
 \left[
 \frac{2}{N}\sum_{J=1}^N \sin (J\theta_m) \cos(J\theta_n) \right] \nn \\
 &\times \left(\frac{ s
 e^{i(\Theta_m-kl)}+e^{-i(\Theta_n-kl)}}{2}\right).
\label{app:vex}
\end{align}
In this equation, we approximate 
$\theta_n \approx 2\pi/3$ or 
$2\sin \theta_n/\sqrt{3} \approx 1$ 
for the states near the Dirac point.
Moreover, $kl$ for those low-energy states 
is a small number
compared with the polar angles $\Theta_n$ and $\Theta_m$.
Thus,  
$se^{i(\Theta_m-kl)}+e^{-i(\Theta_n-kl)}$ approximates
to $e^{i\Theta_m}+e^{-i\Theta_n}$, which is simply
$2\langle \sigma_x \rangle_{mn}$ for $s=1$.
The summation over $J$ in Eq.~(\ref{app:vex}) leads to
\begin{align}
 & \frac{2}{N}\sum_{J=1}^N \sin (J\theta_m) \cos(J\theta_n) \nn \\
 & =
 \begin{cases}
  \displaystyle 0 & \text{$m-n \in$ even} \\
  \displaystyle 
  \frac{2}{\pi} \frac{1}{m-n}+{\cal O}\left(N^{-1}\right)  
  & \text{$m- n \in$ odd}.
 \end{cases}
\label{app:sum}
\end{align}
The correction of ${\cal O}(N^{-1})$ to the matrix element 
has been neglected in the text.
Thus, we reproduce Eqs.~(\ref{eq:ve1}) and (\ref{eq:interx}).
To obtain the right-hand side of Eq.~(\ref{app:sum}),
the quantization of Eq.~(\ref{eq:theta}) is essential.
Namely, the presence of the armchair edge at $J=N$ plays an essential
role.

\section{Derivations of Eqs.~(\ref{eq:vqzig}) and (\ref{eq:vezig})}\label{app:zig}

In this appendix, 
we give the derivation of 
the velocity matrix elements for zigzag nanoribbons.
The electron-light matrix elements are given by
\begin{align}
 & \langle \varphi^{c}_m | \left( -e \bv\cdot {\bf A} \right)  | \varphi^{s}_n \rangle = 
 -ie \frac{\gamma_0}{\hbar}
 \sum_{J=1}^N \left[ \varphi^{c}_{m,J} \right]^\dagger \times \nn \\
 & \left(
 \delta h_{J}^+ \varphi^s_{n,J+1} + 
 \delta h_{J} \varphi^s_{n,J} +
 \delta h_{J}^- \varphi^s_{n,J-1} \right),
 \label{eq:hemzig}
\end{align}
where 
\begin{align}
\begin{split}
 \delta h_{J} & \equiv 
 \sigma_- {\bf A} \cdot \left( e^{-i\frac{ka}{2}} {\bf R}_2 +
 e^{i\frac{ka}{2}} {\bf R}_3 \right) \\
 &- \sigma_+ {\bf A} \cdot 
 \left( e^{-i\frac{ka}{2}} {\bf R}_3 + e^{i\frac{ka}{2}} {\bf R}_2 \right),
 \\
 \delta h_{J}^+ & \equiv \sigma_- {\bf A} \cdot {\bf R}_1,
 \\
 \delta h_{J}^- & \equiv - \sigma_+ {\bf A} \cdot {\bf R}_1.
\end{split}
\end{align}
Putting ${\bf R}_a$ of Eq.~(\ref{app:Ra})
into the above equations,
we get
\begin{align}
\begin{split}
 &\delta h_{J} = 
 \frac{i}{2} \left( 3 \bar{g}(k) A_x \sigma_x + g(k) A_y \sigma_y \right) 
 a_{\rm cc},
 \\
 & \delta h_{J}^+ = \sigma_- A_y a_{\rm cc},
 \\
 & \delta h_{J}^- =- \sigma_+ A_y a_{\rm cc}.
\end{split}
 \label{eq:hzig}
\end{align}
By inserting Eq.~(\ref{eq:hzig}) 
into Eq.~(\ref{eq:hemzig}), we have
\begin{align}
\begin{split}
 & \langle \varphi^{c}_m | v_x  | \varphi^{s}_n \rangle 
 = v_{\rm F}
 \sum_{J=1}^N \left[ \varphi^{c}_{m,J} \right]^\dagger 
 \left\{ -\bar{g}(k)\sigma_x \right\} \varphi^s_{n,J},
 \\
 & \langle \varphi^{c}_m | v_y  | \varphi^{s}_n \rangle 
 = v_{\rm F}
 \sum_{J=1}^N \left[ \varphi^{c}_{m,J} \right]^\dagger 
 \left\{-g(k)\sigma_y+i\frac{2}{3}\varepsilon_n^s \sigma_z
 \right\} \varphi^s_{n,J},
\end{split}
 \label{app:vxezig}
\end{align}
where
$g(k) \equiv 2 \cos\left(ka/2\right)$ and 
$\bar{g}(k) = (2/\sqrt{3}) \sin \left(ka/2\right)$.
Hence, $g=-1$ ($g=1$) for the K (K$'$) point,
and $\bar{g}=1$ for the K and K$'$ points.
To obtain the matrix elements of $v_y$, 
we utilized the energy eigen equation,
\begin{align}
 \sigma_+ \varphi^s_{n,J-1} + K_n^s \varphi^s_{n,J} + \sigma_- \varphi^s_{n,J+1} = 0,
\end{align}
where $K^s_n = g(k) \sigma_x + \varepsilon_n^s$ and $J=1,\ldots,N$.
By multiplying $\sigma_z$ with the eigen equation from the left, we obtain
\begin{align}
 \sigma_+ \varphi^s_{n,J-1} + \sigma_z K^s_n \varphi^s_{n,J} - \sigma_- \varphi^s_{n,J+1} = 0.
\end{align}
Note that the sign in front of $\sigma_-$ becomes minus.
This equation has been used in obtaining the matrix elements of
$v_y$.
In the text,
the term proportional to $\varepsilon^c_n$ in Eq.~(\ref{app:vxezig}) has
been omitted.

\bibliographystyle{apsrev}
%

\end{document}